\documentclass[aps,preprint,superscriptaddress,notitlepage]{revtex4-2} 
\usepackage[utf8]{inputenc}
\usepackage{amsmath, amssymb, bm}       
\usepackage{graphicx}                  
\usepackage{wrapfig}                    
\usepackage{array}                   
\usepackage{booktabs}                  
\usepackage{setspace}                 
\usepackage{placeins}                   
\usepackage{chngcntr}                   
\usepackage{braket}
\usepackage{hyperref}
\usepackage{tikz}                      
\usetikzlibrary{trees}                 

\begin{document}

\title{Experimental demonstration of dual-polarization nonreciprocal thermal emission at moderate magnetic fields}

\author{Mason Christiansen}\affiliation{Department of Materials Science and Engineering, University of California, Los Angeles, Los Angeles, CA 90095, USA}
\author{Jae S. Hwang}\affiliation{Department of Materials Science and Engineering, University of California, Los Angeles, Los Angeles, CA 90095, USA}
\author{Yi Jie Lin}\affiliation{Department of Materials Science and Engineering, University of California, Los Angeles, Los Angeles, CA 90095, USA}
\author{Aryan Zaveri}\affiliation{Department of Materials Science and Engineering, University of California, Los Angeles, Los Angeles, CA 90095, USA}
\author{Tianjun Li}\affiliation{Department of Materials Science and Engineering, University of California, Los Angeles, Los Angeles, CA 90095, USA}
\author{Daniel Cui}\affiliation{Department of Materials Science and Engineering, University of California, Los Angeles, Los Angeles, CA 90095, USA}
\author{Baolai Liang}\affiliation{California NanoSystems Institute, University of California, Los Angeles, Los Angeles, CA 90095, USA}
\author{Aaswath P. Raman}
\email{aaswath@ucla.edu}
\affiliation{Department of Materials Science and Engineering, University of California, Los Angeles, Los Angeles, CA 90095, USA}
\affiliation{California NanoSystems Institute, University of California, Los Angeles, Los Angeles, CA 90095, USA}

\begin{abstract}
Conventional thermal emitters obey Kirchhoff's law of thermal radiation, which requires emissivity and absorptivity to be equal at each wavelength, angle, and polarization. This in turn places a fundamental limit on a range of radiative energy technologies as well as heat transfer. This constraint can however be overcome using magneto-optic semiconductors, which break Lorentz reciprocity under an applied magnetic field. However, in most prior experimental demonstrations, nonreciprocity has been restricted to TM polarization, since the electric field of TE-polarized light is parallel to the applied magnetic field. Here, we experimentally demonstrate nonreciprocal absorption in both polarizations at the same wavelengths and angles from a single structure, wherein a germanium Mie-resonant metasurface on a doped InAs layer scatters TE-polarized light into near fields with components perpendicular to the magnetic field, allowing the magneto-optic response of the InAs to act on them. Under a $0.48\,\mathrm{T}$ field from a permanent Halbach array, this structure exhibits a difference in absorptivity between $+B$ and $-B$ of $0.035$ in both TE and TM polarization at $16.6\,\mu$m. Our approach shows that the magneto-optic response of a planar semiconductor and the resonances of a dielectric metasurface can be designed together, offering a potential route to dual-polarization nonreciprocity at fields available from permanent magnets for emerging energy and heat transfer technologies.
\end{abstract}

\maketitle

\newpage

\section{Introduction}
Thermal radiation plays a central role in infrared imaging, sensing and energy conversion technologies. For a body in thermal equilibrium, Kirchhoff's law of thermal radiation requires the emissivity at a given wavelength, angle and polarization to equal the absorptivity \cite{kirchhoff_ueber_1860}. This equality is a consequence of Lorentz reciprocity and holds for any linear, reciprocal material. Materials that break reciprocity are not subject to this constraint, and their emissivity and absorptivity can be set independently \cite{shayegan_direct_2023,shayegan_broadband_2024,zhang_observation_2025,pajovic_nonreciprocal_2025,wang_maximal_2023}. This independence is useful for energy conversion: nonreciprocal emitters and absorbers have been shown to raise the efficiency of thermophotovoltaic systems and multijunction solar cells \cite{jafari_ghalekohneh_nonreciprocal_2022,park_reaching_2022,park_nonreciprocal_2022,cui_NR_cooling_2026}. The unpolarized nature of thermal and solar radiation however means that these potential gains are only accessible if reciprocity is broken for both TE and TM polarizations.

Several approaches to nonreciprocal thermal emission have been explored. Weyl semimetals have been shown to break time-reversal symmetry without an external field \cite{wu_near-complete_2021,kotov_giant_2018,guo_radiative_2020,zhao_axion-field-enabled_2020}, but work on these materials has so far been largely theoretical. Spatio-temporal modulation of metamaterials offers another route \cite{hadad_breaking_2016,li_nonreciprocity_2022,ghanekar_nonreciprocal_2023}, although the required modulation frequencies, typically in the THz range, are hard to achieve in practice. The strongest experimental violations of Kirchhoff's law so far have come from magneto-optic III-V semiconductors such as GaAs, InAs and InSb \cite{shayegan_direct_2023,shayegan_broadband_2024,zhang_observation_2025,shayegan_nonreciprocal_2022}. Their small bandgaps and light electron effective masses give large magneto-optic responses in the infrared \cite{cui_mo_nr_theory_2026}. In these works, the magnetic field is applied in the Voigt geometry, parallel to the surface and normal to the plane of incidence \cite{shayegan_direct_2023,shayegan_broadband_2024,zhang_observation_2025,shayegan_nonreciprocal_2022}. With this orientation, the electric field of TE-polarized light points along the magnetic field and interacts only with a diagonal component of the permittivity tensor, so TE-polarized emissivity and absorptivity remain equal. Nonreciprocity has therefore been limited to TM polarization, which caps the emissivity--absorptivity contrast for unpolarized light at half of the TM-polarized value. Moreover, the fields used in these experiments, around 1--3~T, are also larger than most applications can supply \cite{shayegan_direct_2023,shayegan_broadband_2024,zhang_observation_2025,shayegan_nonreciprocal_2022}. Theoretical work, however has shown that with effective design, strong violation of Kirchhoff's law below 0.5~T is achievable \cite{zhu_near-complete_2014,zhao_near-complete_2019,cui_mo_nr_theory_2026,jun_nonreciprocity_2025}. 

Several designs have recently been proposed to extend nonreciprocal thermal emission to TE polarization. In a Weyl semimetal film, rotating the plane of incidence away from the Voigt geometry gives TE-polarized light a field component perpendicular to the internal magnetization and makes its absorption nonreciprocal \cite{wu_strong_2023}. For magneto-optic semiconductors under an applied field, silicon nanopore arrays on InAs \cite{fang_dual-polarization_2024} and on a Weyl semimetal \cite{fang_dual-polarization_2024a} are predicted to give strong nonreciprocity in both polarizations through cavity and guided-mode resonances. Additionally a generalized transverse magneto-optic effect has been used to design unpolarized nonreciprocal emitters \cite{zhang_unpolarized_2025}. Metasurfaces on gradient epsilon-near-zero Weyl semimetal multilayers have also been proposed for broadband dual-polarization nonreciprocity \cite{sui_simultaneous_2026}. During the preparation of this manuscript, we became aware of two experimental reports of dual-polarization nonreciprocal absorption, at 1~T \cite{pajovic_dual-polarized_2026} and 1.5~T \cite{xia_dual-polarization_2026}. 

Here, we report the experimental demonstration of nonreciprocal absorption in both TE and TM polarizations using a Mie-resonant dielectric metasurface atop an optimized III-V material platform. The Mie resonances enable strong TE-polarized responses that are designed to align in magnitude with the TM-polarized responses. Under a $\pm0.48\,\mathrm{T}$ magnetic field, we observe a nonreciprocal emissivity--absorptivity contrast of $0.035$ in both polarizations at $16.6\,\mu$m and an angle of $57.3^\circ$. In contrast to recent reports, the TE and TM nonreciprocity we demonstrate here are equal in magnitude at the same wavelength and angle, and the applied field of 0.48~T is less than half of that used in either. Our findings highlight the role of nanophotonic co-design with magneto-optic responses to enable dual-polarization nonreciprocity at moderate magnetic fields accessible with conventional permanent magnets.

\section{Results}

\begin{figure}[htbp]
  \centering
  \includegraphics[width=6.47in]{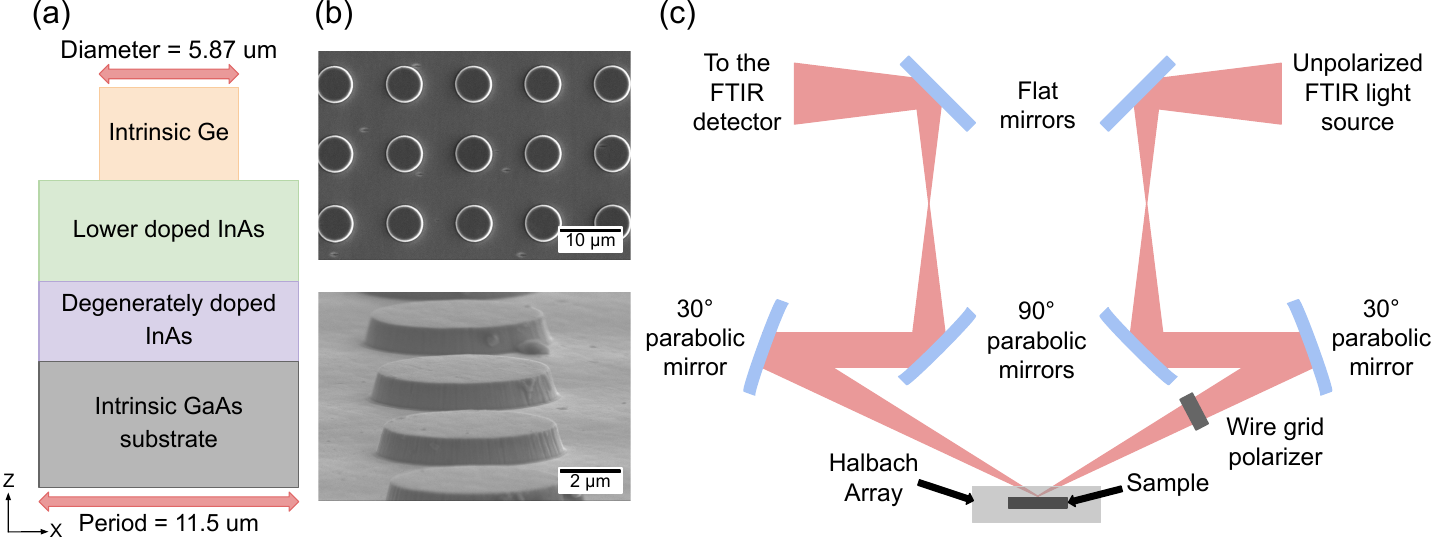}
\caption{(a) Schematic of one period of the sample. An intrinsic GaAs substrate supports a degenerately doped InAs layer that acts as a reflector and a moderately doped InAs layer that provides the magneto-optic response. Germanium pillars of diameter $5.87\,\mu$m on a $11.5\,\mu$m square lattice are deposited on the surface to enable nonreciprocity for TE-polarized light. The $x$ axis lies in the plane of incidence, $z$ is the surface normal, and the magnetic field is applied along $y$. (b) SEM images of the sample. (c) Experimental setup for reflectance measurements as a function of angle under an applied magnetic field. Light from the FTIR source passes through a wire-grid polarizer and is focused onto the sample, which sits at the center of a ring-shaped Halbach array.}
\label{fig1}
\end{figure}

We consider a structure composed of three layers of III-V semiconductors, with germanium pillars deposited on top as shown in Fig. \ref{fig1} (a). The top InAs layer has a dopant concentration of $1.6 \times10^{18}\,\mathrm{cm}^{-3}$ and a thickness of 1 $\mu$m, and is responsible for the nonreciprocal absorptivity and emissivity. The bottom layer of InAs, with a thickness of 250 nm, acts as a reflector in this wavelength range due to its high dopant concentration of $3\times10^{19}\,\mathrm{cm}^{-3}$. The InAs layers are on an intrinsic GaAs substrate. The germanium pillars have a diameter of $5.87\,\mu$m, a height of $\approx$ 1 $\mu$m, and a period of $11.5\,\mu$m, and are what make nonreciprocity in the TE polarization possible. Throughout this paper, $x$ lies in the plane of incidence, $z$ is the surface normal, and the magnetic field $B$ is applied along $y$, the Voigt geometry. The carrier concentration, effective electron mass, and scattering rate of each layer as used for simulation are summarized in Table \ref{tbl1}. SEM images of the fabricated sample can be seen in Fig. \ref{fig1} (b). We measure the reflectance of the sample as a function of wavelength and incident angle under an applied magnetic field using the setup shown in Fig. \ref{fig1} (c), and obtain the absorptivity as $\alpha=1-R$ since the degenerately doped InAs layer blocks transmission.

Nonreciprocity, $\eta$, can be defined as the difference between the absorptivity and emissivity of a material for a given wavelength, angle of incidence, applied magnetic field, and polarization, $\eta = \alpha(\lambda,\theta,B,P)-\mathcal{E}(\lambda,\theta,B,P)$. We know that by the adjoint Kirchhoff's law \cite{Guo_2022} $\alpha(\lambda,\theta,-B,P)=\mathcal{E}(\lambda,\theta,B,P)$. Since the structure is mirror symmetric about the $y$-$z$ plane, reflection across this plane changes the sign of the angle of incidence. Because $B$ is a pseudovector, the same reflection also reverses its $y$ component, which gives $\alpha(\lambda,-\theta,B,P)=\alpha(\lambda,\theta,-B,P)$. Substituting this expression we find that
\begin{align}
\eta=\alpha(\lambda,\theta,B,P)-\alpha(\lambda,\theta,-B,P)
\label{eq:etaB}
\end{align}
which is how we calculate nonreciprocity experimentally. Additionally, we have that
\begin{align}
\eta=\alpha(\lambda,\theta,B,P)-\alpha(\lambda,-\theta,B,P).
\label{eq:etatheta}
\end{align}

The nonreciprocal behavior arises from the magneto-optic response of the active layer, where an applied magnetic field introduces off-diagonal components in the permittivity tensor
\begin{align}
\varepsilon =
\begin{pmatrix}
\varepsilon_{xx} & 0 & \varepsilon_{xz} \\
0 & \varepsilon_{yy} & 0 \\
\varepsilon_{zx} & 0 & \varepsilon_{zz}
\end{pmatrix}
\end{align}
whose elements are defined as
\begin{align}
\varepsilon_{xx} &= \varepsilon_{zz}
= \varepsilon_\infty - \frac{\omega_p^2(\omega + i\gamma)}
{\omega\left[(\omega+i\gamma)^2-\omega_c^2\right]} \\
\varepsilon_{xz} &= -\varepsilon_{zx}
= i \frac{\omega_p^2\omega_c}
{\omega\left[(\omega+i\gamma)^2-\omega_c^2\right]} \\
\varepsilon_{yy} &= \varepsilon_\infty
- \frac{\omega_p^2}{\omega(\omega+i\gamma)}.
\end{align}
Here $\gamma$ is the scattering rate, $\omega_p=\sqrt{n e^2/(m_e \varepsilon_0)}$ is the plasma frequency, $\omega_c = e B/m_e$ is the cyclotron frequency, $n$ is the free carrier concentration, $m_e$ is the electron effective mass, $\varepsilon_\infty$ is the high-frequency permittivity, and $B$ is the applied magnetic field. When $B=0$, $\varepsilon_{xx}=\varepsilon_{yy}=\varepsilon_{zz}$ and the off-diagonal components are zero, so the material is isotropic. When a magnetic field is applied, the permittivity tensor acquires off-diagonal components as the free carriers move in cyclotron motion causing $\varepsilon_{xz}$ and $\varepsilon_{zx}$ to become non-zero. This renders the permittivity tensor asymmetric which break Lorentz reciprocity. For the doping of the active layer, the plasma frequency places the epsilon-near-zero wavelength of $\varepsilon_{yy}$ near $17\,\mu$m, which sets the wavelength of the TM-polarized nonreciprocity \cite{shayegan_broadband_2024}.

Nonreciprocity requires that the free electrons in the active InAs layer be driven by an electric field with a component perpendicular to the applied magnetic field. This in turns causes the cyclotron motion of the carriers, which modifies the material's permittivity function through $\varepsilon_{xz}$ and $\varepsilon_{zx}$. For a planar surface in the Voigt geometry, this condition is met by TM-polarized light, whose electric field has $x$ and $z$ components, but not by TE-polarized light, whose electric field is parallel to $B$ along $y$ and interacts only with $\varepsilon_{yy}$. TE-polarized absorption on a planar InAs surface is therefore the same for $\pm B$, regardless of the doping level or field strength. This motivates the development of the Ge metasurface, which allows for a response in the TE polarization as well. Germanium has a high refractive index and negligible loss in this wavelength range \cite{li_germanium_1980}, so the pillars support Mie resonances whose near fields contain electric field components in all three directions, independent of the polarization of the incident light. 

When TE-polarized light excites these Mie resonances, part of the field is redirected into the $x$ and $z$ directions, and this field extends into the active InAs layer beneath the pillars. Because the scattered field inherits the in-plane wavevector of the incident light, the phase difference between its $x$ and $z$ components depends on the sign of the angle of incidence, and the absorption through $\varepsilon_{xz}$ and $\varepsilon_{zx}$ is therefore different for $\pm\theta$, or equivalently $\pm B$. While the pillars themselves are not magneto-optic, they instead convert TE-polarized light into a field distribution inside the InAs that the magneto-optic response can act on. The pillar diameter, height, and period set the resonance wavelength and the strength of the field in the InAs, which allows the TE-polarized nonreciprocity to be placed near the TM-polarized peak set by the doping of the active layer.

\begin{figure}[htbp]
  \centering
  \includegraphics[width=6.47in]{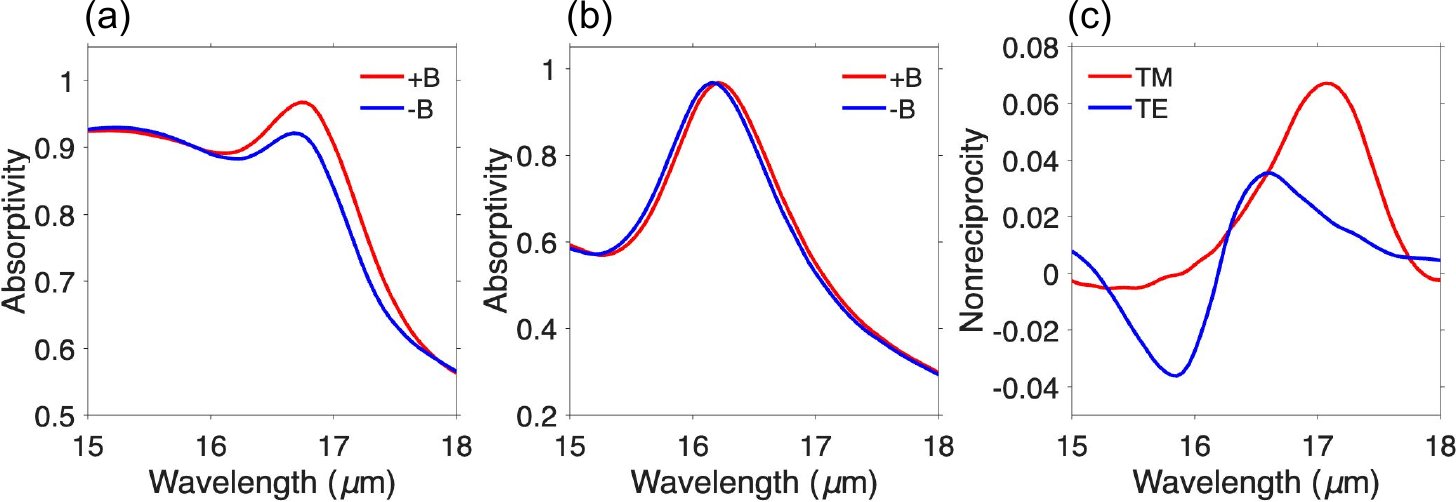}
\caption{Measured absorptivity as a function of wavelength at $\theta=57.3^\circ$ and $B=\pm0.48\,\mathrm{T}$ in the (a) TM and (b) TE polarizations. (c) The difference between the absorptivity at $+B$ and $-B$ for the TM and TE polarizations to show the nonreciprocity of the sample.}
\label{fig2}
\end{figure}

We first measure the absorptivity in the TM and TE polarization for $\pm B$ at $\theta=57.3^\circ$, as shown in Figs. \ref{fig2} (a) and (b). For the TM polarization the absorptivity exceeds $0.88$ from $15\,\mu$m to the peak near $16.7\,\mu$m which reaches $0.96$ at $+B$ and $0.92$ at $-B$, before falling to $0.57$ at $18\,\mu$m as the active layer becomes reflective beyond its plasma wavelength. For the TE polarization, the absorptivity is dominated by the Mie resonance of the pillar array, with a peak of $0.97$ at $16.2\,\mu$m and values of $0.58$ and $0.30$ at the edges of the band. Reversing the field changes the TM spectrum mainly in the height of the peak, while it shifts the TE resonance in wavelength with little change in height; the $+B$ curve lies at slightly longer wavelength than the $-B$ curve. The nonreciprocity in each polarization is simply the difference in absorptivity between $+B$ and $-B$, as shown in Fig. \ref{fig2} (c). For the TM polarization $\eta$ is positive across the band and peaks at $0.067$ near $17.1\,\mu$m. For the TE polarization $\eta$ is antisymmetric about the resonance, reaching $-0.036$ at $15.9\,\mu$m and $+0.035$ at $16.6\,\mu$m, as expected for a field-dependent shift of a resonance. The two curves cross at $16.6\,\mu$m, where both polarizations have $\eta=0.035$, our central result.

\begin{figure}[htbp]
  \centering
  \includegraphics[width=6.47in]{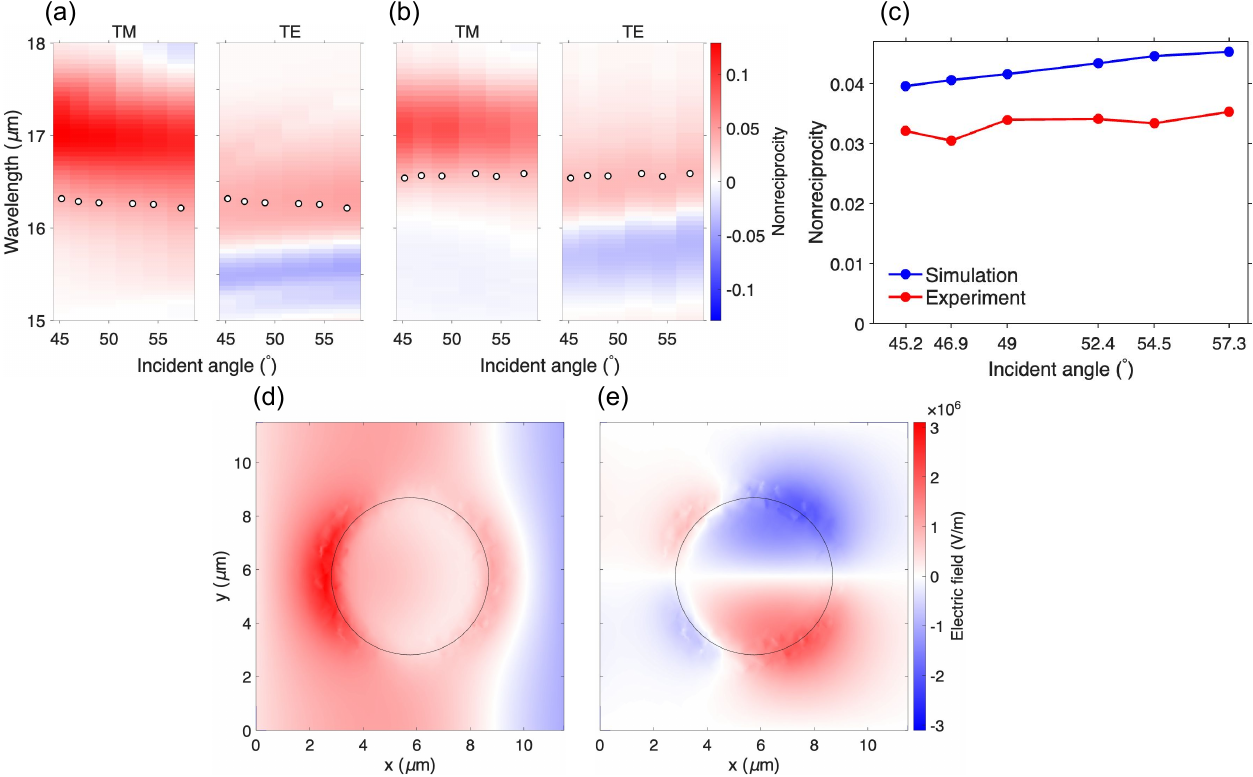}
\caption{(a) Simulated and (b) measured nonreciprocity as a function of wavelength and angle of incidence for TM- and TE-polarized light at $B=0.48\,\mathrm{T}$. The white markers indicate, at each angle, the wavelength at which the nonreciprocity is equal for the two polarizations. These wavelengths lie near the wavelength of peak nonreciprocity in the TE polarization. (c) Nonreciprocity as a function of angle at the wavelengths identified in (a) and (b), comparing simulation and experiment. The simulated real part of the $x$-component of the electric field when $\lambda=16.3\,\mu$m, $\theta=57.3^\circ$, and $B=0.48\,\mathrm{T}$ in the $x$-$y$ plane in the center of the active InAs layer for (d) TM- and (e) TE-polarized light. The circle marks the footprint of the pillar.}
\label{fig3}
\end{figure}

From our simulations we extracted the absorptivity of the sample from $15$ to $18\,\mu$m and from $45.2^\circ$ to $57.3^\circ$ for both TE and TM polarization with an applied magnetic field of $0.48\,\mathrm{T}$, and calculated $\eta$ from Eq. \eqref{eq:etatheta}. The simulated nonreciprocity for the range of angles is shown as color maps for each polarization in Fig. \ref{fig3} (a), alongside the experimentally measured maps in Fig. \ref{fig3} (b). For the TM polarization the simulated nonreciprocity peak occurs near $17\,\mu$m across all angles of incidence and has a peak magnitude of $0.129$. For the TE polarization the simulated nonreciprocity has the same form as in the measurement, negative near $15.6\,\mu$m and positive near $16.3\,\mu$m. The white markers in Fig. \ref{fig3} (a) and (b) show the wavelength at which $\eta$ is equal for the two polarizations. In simulations however this crossing occurs between $16.2$--$16.3\,\mu$m and in the experiments between $16.5$--$16.6\,\mu$m, nearly independent of angle in both cases. Fig. \ref{fig3} (c) plots $\eta$ at the crossing wavelength as a function of angle. The simulated value rises from $0.040$ at $45.2^\circ$ to $0.045$ at $57.3^\circ$, and the measured value lies between $0.031$ and $0.035$ over the same range.

The measured maps reproduce the overall spectral behavior seen in the simulations, with some notable differences. The features in the experiment appear at longer wavelengths than in simulation, with the measured TM peak about half its simulated value ($0.067$ versus $0.129$), and the measured angular dependence at the crossing is flatter. The wavelength offset we observe is consistent with a small difference between the fabricated and designed pillar dimensions, which would shift the Mie resonance's position. Additionally, it is possible that the flatter angular dependence is due to the finite angular spread of the focused FTIR beam, which averages each measurement over a range of incidence angles. Finally, a scattering rate in the active layer higher than the value in Table \ref{tbl1}, dopant diffusion between the InAs layers, and nonuniformity of the field over the illuminated spot, would also result in the same effect. 

To confirm the mechanism described above, we plot in Figs. \ref{fig3} (d) and (e) the simulated real part of $E_x$ in the center of the active InAs layer at the simulated crossing wavelength of $16.3\,\mu$m, $57.3^\circ$, and $0.48\,\mathrm{T}$ for TM- and TE-polarized light. For TM polarization, $E_x$ is present across the whole unit cell because the incident field itself has an $x$ component, and it is enhanced at the edge of the pillar. For TE polarization, the incident field has no $x$ component, and $E_x$ in the InAs appears only through scattering from the pillars; it is concentrated beneath the pillar in two lobes of opposite sign, antisymmetric about the $y$ coordinate of the pillar center, and is comparable in magnitude to the TM case. The sign change does not cancel the nonreciprocity: absorption through $\varepsilon_{xz}$ and $\varepsilon_{zx}$ depends on the local product of $E_x$ and $E_z$, and both components reverse sign together under the mirror symmetry of the TE excitation about the $x$-$z$ plane. The similarity in magnitude of both $E_x$ and $E_z$ between the TM and TE cases is why the two polarizations exhibit similar nonreciprocity near this wavelength.

\section{Conclusion}
We have experimentally demonstrated nonreciprocal thermal emission and absorption in both TE and TM polarizations from a single structure at the same wavelength and angle of incidence. The nonreciprocity originates from the magneto-optic response of a moderately doped InAs layer, with a Ge Mie-resonant metasurface extending this response to TE-polarized light. At an applied field of $0.48\,\mathrm{T}$, provided by a permanent Halbach array, we measured a nonreciprocity of $0.035$ in both polarizations at $16.6\,\mu$m and $57.3^\circ$, with a TM peak of $0.067$ at $17.1\,\mu$m.  We note that the field used here is lower than previous experimental demonstrations, most of which were limited to TM polarization. While our measured nonreciprocity is relatively modest, larger values should be attainable by adjusting the doping to move the TM response closer to the metasurface resonance, and refining the pillar geometry to increase the fraction of TE-polarized field redirected into the InAs. The same approach can be translated to other magneto-optic semiconductors and to other resonant dielectric geometries, allowing for dual-polarized nonreciprocal response across the mid- and far-infrared. Because thermal emission is unpolarized in nature, dual-polarization nonreciprocity is a critical requirement for extending the capabilities of nonreciprocal thermal emitters to real-world applications. Our results here show that it can be achieved with field strengths available from permanent magnets, making nonreciprocal thermal photonics a practical capability for a broad range of energy and heat transfer applications.
\ \\
\ \\
\ \\
\ \\
\ \\
\ \\

\section{Methods}
\subsection{Fabrication}
Molecular beam epitaxy (MBE) was used to grow the InAs layers on a semi-insulating GaAs substrate: a 250 nm layer doped to $3\times10^{19}\,\mathrm{cm}^{-3}$ followed by a 1 $\mu$m layer doped to $1.6\times10^{18}\,\mathrm{cm}^{-3}$. The pillar array was fabricated by first patterning a hole array via photolithography, depositing 0.9 $\mu$m of germanium, and then performing lift-off. The pillar diameter and period measured by SEM are $5.87\,\mu$m and $11.5\,\mu$m respectively.

\subsection{Simulation}
We used the Electromagnetic Waves, Frequency Domain module in COMSOL Multiphysics to perform three-dimensional, full-wave, finite-element simulations of the device. We applied periodic Floquet boundary conditions in the plane of the sample. From the simulation, we calculated the reflectivity of incident light, which was then used to determine the absorptivity ($\alpha(\lambda,\theta)=1-R(\lambda,\theta)$, assuming no transmission). The carrier concentration, effective electron mass, and scattering rate of each layer are summarized in Table \ref{tbl1}, while the Ge pillars were modeled as lossless.

\begin{table}[ht]
\centering
\caption{Material parameters used in simulations}
\setlength{\tabcolsep}{10pt}
\begin{tabular}{lccc}
\hline
\rule{0pt}{12pt}
Layer & Dopant concentration (cm$^{-3}$) & $m_e/m_0$ & $\gamma$ (s$^{-1}$) \\
\hline
\rule{0pt}{12pt}
Active InAs & $1.6\times10^{18}$ & 0.03315 & $5.649\times10^{12}$ \\
Reflector InAs & $3\times10^{19}$ & 0.1299 & $8.881\times10^{12}$ \\
\hline
\end{tabular}
\label{tbl1}
\end{table}

\subsection{Measurement}
In order to apply a magnetic field to the sample, we made the setup shown in Fig. \ref{fig1} (c) which redirected the light from a Bruker Invenio-R FTIR to the sample then back to the FTIR's in-built DTGS detector. A wire-grid polarizer in the incident beam selected TE or TM polarization. By moving the $30^\circ$ parabolic mirrors left or right (symmetrically) and moving the sample and the magnet forwards or backwards, the angle of incidence could be varied, giving incidence angles of $45.2^\circ$, $46.9^\circ$, $49.0^\circ$, $52.4^\circ$, $54.5^\circ$ and $57.3^\circ$. Spectra were referenced to a gold mirror measured at the same angle. The sample sat in the middle of a ring-shaped Halbach array which created a $0.48\,\mathrm{T}$ magnetic field at the position of the sample, in the plane of the sample and perpendicular to the plane of incidence. 

\section{Acknowledgments}

This material is based upon work supported by the National Science Foundation under Grant No.~ECCS-2146577. D.~C. was supported by the National Defense Science and Engineering Graduate (NDSEG) Fellowship. 

\bibliography{bibliography}{}

@article{wu_strong_2023,
  author  = {Wu, J. and Wu, B. and Shi, K. and Wu, X. and Fu, C.},
  title   = {Strong nonreciprocal thermal radiation of transverse electric wave in {Weyl} semimetal},
  journal = {International Journal of Thermal Sciences},
  volume  = {187},
  pages   = {108172},
  year    = {2023},
  doi     = {10.1016/j.ijthermalsci.2023.108172}
}

@misc{cui_NR_cooling_2026,
      title={Nonreciprocal Thermophotonic Cooling}, 
      author={Daniel Cui and Parthiban Santhanam and Aaswath P. Raman},
      year={2026},
      eprint={2604.22814},
      archivePrefix={arXiv},
      primaryClass={physics.app-ph},
      url={https://arxiv.org/abs/2604.22814}, 
}

@article{fang_dual-polarization_2024,
  author  = {Fang, Jimin and Wang, Manzhuo and Liu, Tingyu and Yue, J. and Sun, Xiaoqiang and Wu, Yuanda and Zhang, Daming},
  title   = {Dual-polarization strong nonreciprocal thermal radiation with silicon-based nanopore arrays},
  journal = {International Journal of Thermal Sciences},
  volume  = {195},
  pages   = {108602},
  year    = {2024},
  doi     = {10.1016/j.ijthermalsci.2023.108602}
}

@article{jun_nonreciprocity_2025,
    author = {Wu, Jun and Qing, Ye Ming},
    title = {Mid-infrared strong nonreciprocal thermal radiation with extremely small applied magnetic field},
    journal = {The Journal of Chemical Physics},
    volume = {162},
    number = {11},
    pages = {114706},
    year = {2025},
    month = {03},
    issn = {0021-9606},
    doi = {10.1063/5.0262471},
    url = {https://doi.org/10.1063/5.0262471},
}

@misc{cui_mo_nr_theory_2026,
      title={Magneto-optic perturbation theory for near-complete violation of Kirchhoff's law of thermal emission at low magnetic fields}, 
      author={Daniel Cui and Aaswath P. Raman},
      year={2026},
      eprint={2603.23538},
      archivePrefix={arXiv},
      primaryClass={physics.optics},
      url={https://arxiv.org/abs/2603.23538}, 
}

@article{fang_dual-polarization_2024a,
  author  = {Fang, Jimin and Zou, Jiaqi and Liu, Tingyu and Wang, Manzhuo and Sun, Xiaoqiang and Wu, Yuanda and Zhang, Daming},
  title   = {Dual-polarization small-angle strong nonreciprocal thermal radiator with {Weyl} semimetal},
  journal = {Applied Physics Letters},
  volume  = {124},
  number  = {17},
  pages   = {171702},
  year    = {2024},
  doi     = {10.1063/5.0180575}
}

@article{zhang_unpolarized_2025,
  author  = {Zhang, W. and Li, Z. and Liu, T. and Li, W.},
  title   = {Unpolarized nonreciprocal thermal emitters based on generalized transverse magneto-optical effect},
  journal = {IEEE Journal of Selected Topics in Quantum Electronics},
  year    = {2025},
  doi     = {10.1109/JSTQE.2025.3590100}
}

@article{sui_simultaneous_2026,
  author  = {Sui, Junyang and Liao, Siyuan and Zhang, Haifeng},
  title   = {Simultaneous broadband nonreciprocal thermal emission for transverse magnetic and electric polarizations},
  journal = {Nanophotonics},
  year    = {2026},
  doi     = {10.1002/nap2.70046}
}

@misc{pajovic_dual-polarized_2026,
  author        = {Pajovic, Simo and Zhao, Yiting and Lim, Yae-Chan and Sokhoyan, Ruzan and Atwater, Harry A.},
  title         = {Dual-polarized, mid-infrared nonreciprocal absorption},
  year          = {2026},
  eprint        = {2609.00607},
  archivePrefix = {arXiv},
  primaryClass  = {physics.optics}
}

@misc{xia_dual-polarization_2026,
  author        = {Xia, Shuang and Liu, Mengqi and Wan, Wenjian and Wang, Jialong and Yang, Weihao and Wang, Chaoran and Ma, Huiqin and Qin, Jun and Li, Hua and Wang, Yuan and Bi, Lei and Qiu, Cheng-Wei and Yin, Xiaobo},
  title         = {Dual-polarization control of broadband nonreciprocal thermal radiation by combining local and nonlocal metasurfaces},
  year          = {2026},
  eprint        = {2608.05640},
  archivePrefix = {arXiv}
}

@article{li_germanium_1980,
    author = {Li, H. H.},
    title = {Refractive index of silicon and germanium and its wavelength and temperature derivatives},
    journal = {Journal of Physical and Chemical Reference Data},
    volume = {9},
    number = {3},
    pages = {561-658},
    year = {1980},
    month = {07},
    issn = {0047-2689},
    doi = {10.1063/1.555624},
    url = {https://doi.org/10.1063/1.555624}
}

@article{wang_maximal_2023,
  title = {Maximal violation of Kirchhoff's law in planar heterostructures},
  volume = {5},
  issn = {2643-1564},
  url = {https://link.aps.org/doi/10.1103/PhysRevResearch.5.L022051},
  doi = {10.1103/PhysRevResearch.5.L022051},
  pages = {L022051},
  number = {2},
  journal = {Physical Review Research},
  shortjournal = {Phys. Rev. Research},
  author = {Wang, Lu and García De Abajo, F. Javier and Papadakis, Georgia T.},
  urldate = {2025-03-14},
  date = {2023-06-07},
  year = {2023},
  langid = {english},
}

@article{pajovic_nonreciprocal_2025,
  title = {Nonreciprocal reflection of mid-infrared light by highly doped {InAs} at low magnetic fields},
  volume = {33},
  issn = {1094-4087},
  url = {https://opg.optica.org/abstract.cfm?URI=oe-33-4-8661},
  doi = {10.1364/OE.544496},
  pages = {8661},
  number = {4},
  journal = {Optics Express},
  shortjournal = {Opt. Express},
  author = {Pajovic, Simo and Tsurimaki, Yoichiro and Qian, Xin and Chen, Gang and Boriskina, Svetlana V.},
  urldate = {2025-09-18},
  date = {2025-02-24},
  year = {2025},
  langid = {english},
}

@article{zhao_near-complete_2019,
  title = {Near-complete violation of Kirchhoff’s law of thermal radiation with a 03 T magnetic field},
  volume = {44},
  issn = {0146-9592, 1539-4794},
  url = {https://opg.optica.org/abstract.cfm?URI=ol-44-17-4203},
  doi = {10.1364/OL.44.004203},
  pages = {4203},
  number = {17},
  journal = {Optics Letters},
  shortjournal = {Opt. Lett.},
  author = {Zhao, Bo and Shi, Yu and Wang, Jiahui and Zhao, Zhexin and Zhao, Nathan and Fan, Shanhui},
  urldate = {2025-09-18},
  date = {2019-09-01},
  year = {2019},
  langid = {english},
}

@article{shayegan_direct_2023,
  title = {Direct observation of the violation of Kirchhoff’s law of thermal radiation},
  volume = {17},
  issn = {1749-4885, 1749-4893},
  url = {https://www.nature.com/articles/s41566-023-01261-6},
  doi = {10.1038/s41566-023-01261-6},
  pages = {891--896},
  number = {10},
  journal = {Nature Photonics},
  shortjournal = {Nat. Photon.},
  author = {Shayegan, Komron J. and Biswas, Souvik and Zhao, Bo and Fan, Shanhui and Atwater, Harry A.},
  urldate = {2025-09-18},
  date = {2023-10},
  year = {2023},
  langid = {english}
}

@article{shayegan_broadband_2024,
  title = {Broadband nonreciprocal thermal emissivity and absorptivity},
  volume = {13},
  issn = {2047-7538},
  url = {https://www.nature.com/articles/s41377-024-01520-3},
  doi = {10.1038/s41377-024-01520-3},
  pages = {176},
  number = {1},
  journal = {Light: Science \& Applications},
  shortjournal = {Light Sci Appl},
  author = {Shayegan, Komron J. and Hwang, Jae S. and Zhao, Bo and Raman, Aaswath P. and Atwater, Harry A.},
  urldate = {2025-09-18},
  date = {2024-07-24},
  year = {2024},
  langid = {english},
}

@article{zhu_near-complete_2014,
  title = {Near-complete violation of detailed balance in thermal radiation},
  volume = {90},
  rights = {http://link.aps.org/licenses/aps-default-license},
  issn = {1098-0121, 1550-235X},
  url = {https://link.aps.org/doi/10.1103/PhysRevB.90.220301},
  doi = {10.1103/PhysRevB.90.220301},
  pages = {220301},
  number = {22},
  journal = {Physical Review B},
  shortjournal = {Phys. Rev. B},
  author = {Zhu, Linxiao and Fan, Shanhui},
  urldate = {2025-09-18},
  date = {2014-12-01},
  year = {2014},
  langid = {english},
}

@article{kirchhoff_ueber_1860,
  title = {Ueber das Verhältniss zwischen dem Emissionsvermögen und dem Absorptionsvermögen der Körper für Wärme und Licht},
  volume = {185},
  issn = {0003-3804, 1521-3889},
  url = {https://onlinelibrary.wiley.com/doi/10.1002/andp.18601850205},
  doi = {10.1002/andp.18601850205},
  pages = {275--301},
  number = {2},
  journal = {Annalen der Physik},
  shortjournal = {Annalen der Physik},
  author = {Kirchhoff, G.},
  urldate = {2025-09-18},
  date = {1860-01},
  year = {1860},
  langid = {german},
}

@article{jafari_ghalekohneh_nonreciprocal_2022,
  title = {Nonreciprocal Solar Thermophotovoltaics},
  volume = {18},
  issn = {2331-7019},
  url = {https://link.aps.org/doi/10.1103/PhysRevApplied.18.034083},
  doi = {10.1103/PhysRevApplied.18.034083},
  pages = {034083},
  number = {3},
  journal = {Physical Review Applied},
  shortjournal = {Phys. Rev. Applied},
  author = {Jafari Ghalekohneh, Sina and Zhao, Bo},
  urldate = {2025-09-18},
  date = {2022-09-29},
  year = {2022},
  langid = {english},
}

@article{park_reaching_2022,
  title = {Reaching the Ultimate Efficiency of Solar Energy Harvesting with a Nonreciprocal Multijunction Solar Cell},
  volume = {22},
  rights = {https://doi.org/10.15223/policy-029},
  issn = {1530-6984, 1530-6992},
  url = {https://pubs.acs.org/doi/10.1021/acs.nanolett.1c04288},
  doi = {10.1021/acs.nanolett.1c04288},
  pages = {448--452},
  number = {1},
  journal = {Nano Letters},
  shortjournal = {Nano Lett.},
  author = {Park, Yubin and Zhao, Bo and Fan, Shanhui},
  urldate = {2025-09-18},
  date = {2022-01-12},
  year = {2022},
  langid = {english},
}

@article{kotov_giant_2018,
  title = {Giant tunable nonreciprocity of light in Weyl semimetals},
  volume = {98},
  issn = {2469-9950, 2469-9969},
  url = {https://link.aps.org/doi/10.1103/PhysRevB.98.195446},
  doi = {10.1103/PhysRevB.98.195446},
  pages = {195446},
  number = {19},
  journal = {Physical Review B},
  shortjournal = {Phys. Rev. B},
  author = {Kotov, O. V. and Lozovik, Yu. E.},
  urldate = {2025-09-18},
  date = {2018-11-30},
  year = {2018},
  langid = {english},
}

@article{zhao_axion-field-enabled_2020,
  title = {Axion-Field-Enabled Nonreciprocal Thermal Radiation in Weyl Semimetals},
  volume = {20},
  rights = {https://doi.org/10.15223/policy-029},
  issn = {1530-6984, 1530-6992},
  url = {https://pubs.acs.org/doi/10.1021/acs.nanolett.9b05179},
  doi = {10.1021/acs.nanolett.9b05179},
  pages = {1923--1927},
  number = {3},
  journal = {Nano Letters},
  shortjournal = {Nano Lett.},
  author = {Zhao, Bo and Guo, Cheng and Garcia, Christina A. C. and Narang, Prineha and Fan, Shanhui},
  urldate = {2025-09-18},
  date = {2020-03-11},
  year = {2020},
  langid = {english},
}

@article{guo_radiative_2020,
  title = {Radiative Thermal Router Based on Tunable Magnetic Weyl Semimetals},
  volume = {7},
  rights = {https://doi.org/10.15223/policy-029},
  issn = {2330-4022, 2330-4022},
  url = {https://pubs.acs.org/doi/10.1021/acsphotonics.0c01376},
  doi = {10.1021/acsphotonics.0c01376},
  pages = {3257--3263},
  number = {11},
  journal = {ACS Photonics},
  shortjournal = {ACS Photonics},
  author = {Guo, Cheng and Zhao, Bo and Huang, Danhong and Fan, Shanhui},
  urldate = {2025-09-18},
  date = {2020-11-18},
  year = {2020},
  langid = {english},
}

@article{wu_near-complete_2021,
  title = {Near-complete violation of Kirchhoff’s law of thermal radiation in ultrathin magnetic Weyl semimetal films},
  volume = {11},
  issn = {2159-3930},
  url = {https://opg.optica.org/abstract.cfm?URI=ome-11-12-4058},
  doi = {10.1364/OME.444308},
  pages = {4058},
  number = {12},
  journal = {Optical Materials Express},
  shortjournal = {Opt. Mater. Express},
  author = {Wu, Jun and Wang, Zhongmin and Zhai, Han and Shi, Zhangxing and Wu, Xiaohu and Wu, Feng},
  urldate = {2025-09-18},
  date = {2021-12-01},
  year = {2021},
  langid = {english},
}

@article{park_nonreciprocal_2022,
  title = {Nonreciprocal Thermophotovoltaic Systems},
  volume = {9},
  rights = {https://doi.org/10.15223/policy-029},
  issn = {2330-4022, 2330-4022},
  url = {https://pubs.acs.org/doi/10.1021/acsphotonics.2c01308},
  doi = {10.1021/acsphotonics.2c01308},
  pages = {3943--3949},
  number = {12},
  journal = {ACS Photonics},
  shortjournal = {ACS Photonics},
  author = {Park, Yubin and Omair, Zunaid and Fan, Shanhui},
  urldate = {2025-09-18},
  date = {2022-12-21},
  year = {2022},
  langid = {english},
}

@article{zhang_observation_2025,
  title = {Observation of Strong Nonreciprocal Thermal Emission},
  volume = {135},
  issn = {0031-9007, 1079-7114},
  url = {https://link.aps.org/doi/10.1103/PhysRevLett.135.016901},
  doi = {10.1103/PhysRevLett.135.016901},
  pages = {016901},
  number = {1},
  journal = {Physical Review Letters},
  shortjournal = {Phys. Rev. Lett.},
  author = {Zhang, Zhenong and Kalantari Dehaghi, Alireza and Ghosh, Pramit and Zhu, Linxiao},
  urldate = {2025-09-18},
  date = {2025-06-30},
  year = {2025},
  langid = {english},
}

@article{ghanekar_nonreciprocal_2023,
  title = {Nonreciprocal Thermal Emission Using Spatiotemporal Modulation of Graphene},
  volume = {10},
  rights = {https://doi.org/10.15223/policy-029},
  issn = {2330-4022, 2330-4022},
  url = {https://pubs.acs.org/doi/10.1021/acsphotonics.2c01411},
  doi = {10.1021/acsphotonics.2c01411},
  pages = {170--178},
  number = {1},
  journal = {ACS Photonics},
  shortjournal = {ACS Photonics},
  author = {Ghanekar, Alok and Wang, Jiahui and Guo, Cheng and Fan, Shanhui and Povinelli, Michelle L.},
  urldate = {2025-09-18},
  date = {2023-01-18},
  year = {2023},
  langid = {english},
}

@article{hadad_breaking_2016,
  title = {Breaking temporal symmetries for emission and absorption},
  volume = {113},
  issn = {0027-8424, 1091-6490},
  url = {https://pnas.org/doi/full/10.1073/pnas.1517363113},
  doi = {10.1073/pnas.1517363113},
  pages = {3471--3475},
  number = {13},
  journal = {Proceedings of the National Academy of Sciences},
  shortjournal = {Proc. Natl. Acad. Sci. U.S.A.},
  author = {Hadad, Yakir and Soric, Jason C. and Alu, Andrea},
  urldate = {2025-09-18},
  date = {2016-03-29},
  year = {2016},
  langid = {english},
}

@article{shayegan_nonreciprocal_2022,
  title = {Nonreciprocal infrared absorption via resonant magneto-optical coupling to {InAs}},
  volume = {8},
  issn = {2375-2548},
  url = {https://www.science.org/doi/10.1126/sciadv.abm4308},
  doi = {10.1126/sciadv.abm4308},
  pages = {eabm4308},
  number = {18},
  journal = {Science Advances},
  shortjournal = {Sci. Adv.},
  author = {Shayegan, Komron J. and Zhao, Bo and Kim, Yonghwi and Fan, Shanhui and Atwater, Harry A.},
  urldate = {2025-09-18},
  date = {2022-05-06},
  year = {2022},
  langid = {english},
}

@article{li_nonreciprocity_2022,
  title = {Nonreciprocity and Faraday Rotation at Time Interfaces},
  volume = {128},
  issn = {0031-9007, 1079-7114},
  url = {https://link.aps.org/doi/10.1103/PhysRevLett.128.173901},
  doi = {10.1103/PhysRevLett.128.173901},
  pages = {173901},
  number = {17},
  journal = {Physical Review Letters},
  shortjournal = {Phys. Rev. Lett.},
  author = {Li, Huanan and Yin, Shixiong and Alù, Andrea},
  urldate = {2025-09-21},
  date = {2022-04-29},
  year = {2022},
  langid = {english},
}

@article{Guo_2022, title={Adjoint Kirchhoff’s Law and General Symmetry Implications for All Thermal Emitters}, volume={12}, DOI={10.1103/PhysRevX.12.021023}, number={2}, journal={Physical Review X}, author={Guo, Cheng}, year={2022} }
\end{document}